\documentclass[prc,aps,eqsecnum,amssymb,floatfix,showpacs]{revtex4}
\usepackage{graphicx}% Include figure files
\usepackage{dcolumn}% Align table columns on decimal point
\usepackage{bm}% bold math
\usepackage[english]{babel}
\usepackage{longtable}
\usepackage[inline]{trackchanges}
\addeditor{sanjay}
\addeditor{joe}
\addeditor{gang}

\def\gsim{\mathrel{\rlap{\lower4pt\hbox{\hskip1pt$\sim$}}}}
\newcommand{\beq}{\begin{equation}}
\newcommand{\eeq}{\end{equation}}
\newcommand{\beqn}{\begin{eqnarray}}
\newcommand{\eeqn}{\end{eqnarray}}
\newcommand{\btab}{\begin{tabular}}
\newcommand{\etab}{\end{tabular}}

\def\ms{M_s}

\newcommand{\kk}{{\bf k}}

\begin{document}
{\begin{flushleft} INT-PUB-12-027 \\
LA-UR-12-21609
\vskip -0.5in
{\normalsize }
\end{flushleft}

\title{Spin Response and Neutrino Emissivity of Dense Neutron Matter}

\author{G.\ Shen$^{\,{\rm a}}$, S.\ Gandolfi$^{\,{\rm a}}$, S.\ Reddy $^{\,{\rm a,b}}$, and J.\ Carlson$^{\,{\rm a}}$}

\affiliation{
$^{\rm a}$\mbox{Theoretical Division, Los Alamos National Laboratory, NM 87545, USA}\\
$^{\rm b}$\mbox{Institute for Nuclear Theory, University of Washington, Seattle, WA 98195, USA}\\
}
\date{\today}

\begin{abstract}

We study the spin response of cold dense neutron matter in the limit of zero momentum transfer, and show that the frequency
dependence of the long-wavelength spin response is well constrained by sum-rules
and the asymptotic behavior of the two-particle response at high frequency.  The
sum-rules are calculated using Auxiliary
Field Diffusion Monte Carlo technique and the high frequency two-particle 
response is calculated for several nucleon-nucleon potentials. At nuclear saturation density,
the sum-rules suggest that the strength of the spin response peaks at $\omega
\simeq$ 40--60 MeV, decays rapidly for $\omega \geq $100 MeV, and has a sizable
strength below 40 MeV.  This strength at relatively low energy may lead to 
enhanced neutrino production rates in dense neutron-rich matter at temperatures of relevance to core-collapse supernova. 
\end{abstract}

\maketitle
%%%%%%%%%%%%%%%%%%%%%%%%% introduction %%%%%%%%%%%%%%%%%%%%%%%

%%%% neutrino response and spin response in SN
\section{Introduction}
The spin response of dense neutron matter plays an essential role in
determining neutrino interaction rates in neutron stars and 
supernovae~\cite{Friman:1979,Iwamoto:1982,Raffelt:1996,Reddy:1998hb,Lykasov:2008}. (For the effect of spin response on photon interaction with nucleon magnetic moments, see discussion in Ref.~\cite{Kopf:1998}.)
Since the energy and momentum transfer between neutrinos and matter is small compared to 
the Fermi energy and momentum, degeneracy and many body effects can strongly modify interaction rates.   
The spin response of neutron matter is an intriguing problem in that a non-zero response
requires the coupling of spin and space through the tensor and spin-orbit components of the nuclear
force.

We study the response in the specific limit of zero temperature and zero momentum transfer, and discuss how this limiting case will be useful to understand the more general behavior encountered at finite temperatures in neutron stars and supernovas.  At zero temperature the spin response can be obtained through a combination of 
sum-rules and a calculation of the high-energy part of the
response.  The sum-rules and the high-energy behavior resolve nuclear interactions with momenta
of the order of the pion wavelength, and we use nuclear Hamiltonians previously found to be reliable in describing relevant excitations and their coupling to the ground state in the other contexts.

Bremsstrahlung reactions such as  $n+n \rightarrow n+n+\nu+\bar{\nu}$ are an important source of neutrino pair production in dilute neutron matter. When neutrons are non-relativistic, neutrino emission occurs primarily 
due to fluctuations of the nucleon spin. Density and current fluctuations are suppressed by the
velocity due to particle number and momentum conservation. This dominance of spin
fluctuations is unique feature of nuclear systems because strong non-central
tensor and spin-orbit forces that do not commute with the spin operator lead to
enhanced spin fluctuations even in the long-wavelength limit ($\bm q\rightarrow
0$). Its importance in neutrino production rates was first realized in
pioneering work by Friman and Maxwell \cite{Friman:1979}. They calculated the
neutrino production rate in the long-wavelength limit using the one-pion-exchange
(OPE)  potential in leading order perturbation theory (Born approximation). In
subsequent work, Hanhart et al. calculated the neutrino production rate in
neutron matter using a low-energy approximation that relates the rate directly
to observed nucleon-nucleon phase shifts obviating the need to rely on either
perturbation theory or a specific choice for the nucleon-nucleon potential
\cite{Hanhart:2001}. While these calculations have provided a useful benchmark, they neglect many-body effects and their regime of validity is
restricted to weakly correlated dilute systems. 

The inclusion of  many-body effects have relied on diagrammatic perturbation theory 
where specific corrections to long-distance and long-time behavior of nucleon propagation 
in the medium are incorporated. The finite lifetime of 
quasi-particles, screening of the weak axial charge, as well as screening of
nucleon-nucleon interactions due to particle-hole polarization effects at finite
density have been investigated by several authors \cite{Sigl:1996,Hannestad:1998,vanDalen:2003,Schwenk:2004,Lykasov:2005,Lykasov:2008}. These calculations have shown that these corrections are important and generically
tend to decrease neutrino production rates.  
On the other hand, attempts to include in-medium softening of the pion propagator and corrections to the nucleon propagators and weak vertices \cite{Voskresensky:1986,Migdal:1990,Voskresensky:2001}  have shown that the neutrino emissivity can be enhanced. 
However, these methods neglect terms in many-body perturbation theory and 
it is presently difficult to estimate associated errors.  To overcome this shortcoming, 
we adopt a different strategy, and use Quantum Monte Carlo (QMC) to compute three lowest order 
sum-rules which are described below in \S\ref{sec:sumrules}.  We supplement these 
sum-rules with asymptotic form of the two-particle response valid at high frequency to 
deduce the distribution of strength of the spin response function at lower energies of relevance to astrophysics.

\section{Neutrino Emissivity and the Spin Structure Function}
\label{sec:basic} 

From the point of view of many-body theory, neutrino interaction rates in the
medium can be factored into a product of two terms: (i) the correlation functions of the
dense medium, and (ii) kinematical factors and coupling constants associated
with neutrino currents. The latter are well-known and relatively simple functions of the neutrino energy and momentum. In contrast, the spin, density and current correlation functions are complex functions of temperature, density, and the energy and momentum transfer because multi-particle dynamics and correlations in the ground state of the strongly interacting system play a critical role. 

The dynamic spin structure factor $S_\sigma(\omega,{\bf q})$ of neutron matter
encodes the linear response of neutron matter to spin fluctuations and
is defined as \cite{Iwamoto:1982} 
\beqn
\label{SA} 
S_\sigma(\omega,{\bf q})\ &=&\ \frac {4} {3n} \frac{1}{2\pi} \int_{-\infty}^{\infty} dt 
e^{i\omega t} \bigl < {\bf s}(t,{\bf q})\cdot{\bf s}(0,{\bf -q}) \bigr > \, \nonumber \\ 
& = & \frac{4}{3n} \sum_f \langle 0 | s({\bf q}) | f \rangle  \cdot \langle f | s({\bf -q}) | 0 \rangle \delta (\omega - (E_f - E_0)) \,
\eeqn
where ${\bf s}(t,q)=V^{-1}\sum_{i=1}^N e^{-i{\bm q}\cdot {\bm r_i}(t)}\bm
\sigma_i$ and $\bm \sigma_i$ is the spin operator acting on the $i^{th}$ nucleon
at time $t$. The second line expresses the same response as a sum over final states $| f \rangle$ coupled to the ground state through the time-independent spin operator.

Alternatively in terms of the field operators, ${\bf s}(t,q)$ is
Fourier transform of the spin density operator ${\bf s}(x) =\frac{1}{2}\psi^+(x)
~{\bm \tau}~ \psi(x) $ with ${\bm \tau}$ being the usual Pauli matrix and
$\psi(x)$ is the non-relativistic field operator. The normalization factor  $4/
3n$ where $n$ is the neutron number density ensures that the dynamic form factor
is canonically normalized such that $S(q \rightarrow \infty) =1$ for the non-interacting Fermi systems and 
conforms to the standard definitions of the sum-rules discussed below in \S\ref{sec:sumrules}.   

The rate of neutrino pair production can 
be expanded in powers of the nucleon velocity and the momentum of the neutrino pair  \cite{Friman:1979}.  
The neutrino emissivity of neutron matter denoted by $Q$, and defined as the rate of energy loss due to
neutrino pair production per unit volume and per unit time, to leading order in the neutron velocity and neutrino momentum 
is given by\cite{Hannestad:1998}    
\beq
\label{eloss} 
Q\ =\ \frac{C_A^2G_F^2 n}{20\pi^3}\int_0^{\infty} d\omega~ \omega^6 ~e^{-\omega/T} S_\sigma(\omega) \,,
\eeq
where $G_F=1.18\times10^{-11}$ MeV$^{-2}$ is the Fermi
constant of the weak interaction, $C_A=-1.26/2$ is the neutron neutral-current axial coupling constant. Note due to the strange-quark contribution to the nucleon spin, $C_A$ may be modified in neutral current processes by few percent in the energy range of interest to supernova \cite{Raffelt:1995}. Here we have not included this modification for simplicity.
At low temperature, when $T \ll E_{\rm Fn}$, where $E_{\rm Fn}$ is the neutron Fermi energy, the neutrino pair momentum ${\bf q}$ is small compared to the both the Fermi momentum $k_{Fn}$ and the intrinsic momentum scales associated with the strong interaction, and may be neglected and $S_\sigma(\omega)= S_\sigma(\omega,{\bf q}=0)$. Hence in Eq.~(\ref{eloss}) only $S_\sigma(\omega)$ appears and it is both a function of density and temperature as implied by the ensemble average denoted on the RHS of the equation.

\section{sum-rules}
\label{sec:sumrules}

The spin response describes the coupling to the ensemble of
final states obtained by flipping all the ground-state spins in 
neutron matter.  If spin and space are uncoupled, spin is a good
quantum number and there would be no response at zero momentum transfer.
However, the spin-orbit and tensor interactions (acting only in 
relative $p-$waves and higher in neutron matter) induce a finite expectation
value of $\langle S^2 \rangle$ even at $T$=0 and a finite response results.
The spin-orbit and tensor interactions are of pion range or less, so they 
predominantly affect neutrons coupled to spin 1 at a pair separation typical for 
nearest neighbors at that density. Although there is zero total momentum transfer, 
the two interacting particles can nevertheless have significant relative
momenta in the relevant final states.

The overall strength and energy distribution of the response can be
characterized through the relevant sum-rules.  We 
employ QMC to compute the low order sum-rules that relate
moments of $S_\sigma(\omega,{\bf q})$ to its ground state properties. 
We then combine these sum-rule constraints with asymptotic high-energy behavior 
expected in the two-particle system to obtain constraints on the distribution of 
strength of $S_\sigma(\omega)$ as a function of $\omega$ at $\bm q=0$.   
For the same reason, the
response in Eq.~(\ref{SA}) is solely due to the excitation of multi-particle
states as single particle excitations vanish for these kinematics.   

Though we ultimately desire information about the spectrum and coupling to
the excited states of the system, the moments of the sum-rules
defined by the relation 
\beq
\label{SUM}
S^n_\sigma  =  \int_0^\infty S_\sigma (\omega,\bm q=0)~\omega^n~ d\omega, 
\eeq
are calculable as ground state properties.  The sum-rules provide a simple and
systematic means to eliminate explicit dependence on the intermediate excited
states of the system.The
relevant excited state information is sampled by
operators  contained in the nuclear Hamiltonian. In this study we use the
following sum-rule relations:
 \beqn
 \label{S1}
 S^{-1}_\sigma &=& \frac{\chi_\sigma}{2n} \, \\
%= \frac{\chi_\sigma^F}{2n(1+G_0)} \, \\
 S^{0}_\sigma &=& \ 1\ +\ \lim_{q \rightarrow 0}
\label{S2}
\frac{4}{3N}\sum\limits_{i\ne j}^N\bigl < 0 | e^{-i{\bf q}\cdot (\bf r_i
- r_j)} {\boldsymbol\sigma_i}\cdot {\boldsymbol\sigma_j} | 0  \bigr > \,\\
\label{S3}
S^{+1}_\sigma &=& -\frac{4}{3N} \  \lim_{q \rightarrow 0}
\bigl < 0 |[H_N,{\boldsymbol s}({\bf q})] \cdot 
{\boldsymbol s(-{\bf q})}| 0  \bigr > \, 
 \eeqn 
where $\chi_\sigma= \partial n_\sigma /\partial \mu_\sigma$ is the spin
susceptibility of the interacting ground state $| 0 \bigr >$ of the nuclear
Hamiltonian $H_N$, and $n_\sigma$ and $ \mu_\sigma$ are number density and chemical potential of particles with spin $\sigma$ ($\pm1/2$). 
%In Eq.~(\ref{S1}),
%the spin susceptibility of the interacting
%system 
%is related to the Pauli susceptibility of the ideal Fermi gas 
%$ \chi_\sigma^F=
%N(0)$ where $N(0)=m k_F/\pi^2$ is the density of states at the Fermi surface,
%and the Landau parameter $G_0$ characterizes the non-ideal behavior.  
Our
strategy here is to evaluate the right hand side of Eqs.~(\ref{S1}), (\ref{S2}) and (\ref{S3})
using QMC and use this information to constrain the behavior of $S(\omega)$
for values of $\omega$ relevant to the calculation of neutrino production.  

This strategy is not new, in
Ref.~\cite{Sigl:1996} estimates of the $S^0_\sigma$ and $S^1_\sigma$
sum-rules were used to argue that spin response function must saturate at high
density, and in Ref.~\cite{Olsson:2004}, sum-rules were used to estimate the
relative importance of multi-particle excitations to the response function in
the kinematical regime where $\omega \ge q$. Our work improves upon these
earlier studies in two respects: (i) we compute and combine for the first time
all three sum-rules to constrain both low-frequency and high-frequency behavior
of  $S(\omega,\bm q=0)$; and (ii) we deduce the high-frequency response or
short-time behavior of the two-particle dynamics where they
dominate in the many-body
system by direct calculation of the two-particle matrix elements. 

%%%% variational and QMC calculations of response, pairing is not present at high T
To compute the expectation values of operators in the ground state needed to
evaluate the sum-rules we use a non-relativistic nuclear Hamiltonian with local
2-body potentials of the form    \beq  
\label{Ham}
%H = \sum\limits_i^N \frac{P_i^2}{2m} +\ 
%\frac 1 2 \sum\limits_{i\neq j}^N V(r_{ij}, \sigma_1,\sigma_2). 
H_N = \sum\limits_i^N \frac{{\bf p}_i^2}{2m} +\ 
\sum\limits_{i<j} \sum\limits_{p}^4 v_p(r_{ij})O^{(p)}_{ij}. 
\eeq 
where  $O^{p=1,4}_{ij}=(1,\boldsymbol\sigma_i\cdot\boldsymbol\sigma_j,
S_{ij},\boldsymbol L\cdot \boldsymbol S)$, and $S_{ij}=(3\boldsymbol\sigma_1\cdot \hat{r}\boldsymbol\sigma_2\cdot \hat{r} 
-\boldsymbol\sigma_1\cdot\boldsymbol\sigma_2)$ 
is the tensor operator, and $\boldsymbol L\cdot \boldsymbol S$ is the spin-orbit
operator. We employ the Auxiliary Field Diffusion Monte Carlo (AFDMC) method
described in Ref.~\cite{Schmidt:1999,Gandolfi:2009} and use the Argonne AV8'
form for the two-body interaction as it provides a good description of
properties of light nuclei~\cite{Pudliner:1997}.  
The AFDMC calculations use auxiliary field quantum Monte Carlo techniques to
treat the spin and spatial degrees of freedom in neutron matter. They
have been used extensively to calculate the equation of state of neutron matter,
and also the spin susceptibility \cite{Fantoni:2001}. 
We use AFDMC to compute the sum-rules expressed in Eq. (\ref{S1}, \ref{S2})
and (\ref{S3}). Note the static structure function $S_\sigma^0$ and energy-weighted sum rule $S_\sigma^1$ have been previously evaluated for Argonne potentials~\cite{Wiringa:1995}, first based on variational methods \cite{Akmal:1997,Cowell:2004}.

The $S_\sigma^{-1}$ sum-rule is calculated by considering the ground state
of neutron matter in the presence of a magnetic field as proposed in
Ref.~\cite{Fantoni:2001}. 
The energy of neutron matter in the presence of a magnetic field is:
\begin{equation}
E (p)\ = E (0) - b P + (1/2) P^2 E''(0),
\end{equation}
where $E(0)$ is the ground state energy in the absence of a magnetic field,
$P = \frac{N_\uparrow - N_\downarrow} { N_\uparrow + N_\downarrow}$ is the spin polarization, and the spin susceptibility $\chi_\sigma$ is
\begin{equation}
\chi_\sigma \ = \ \mu^2 P \frac{1}{E'' (0)}.
\end{equation}
The calculations are performed for zero magnetic field and a finite
magnetic field for of order 60 particles in periodic boundary conditions.
The system we simulate has finite number of up and down neutrons, and
the magnetic field is chosen in such a way the finite system is close to
the thermodynamic limit as described in Ref.~\cite{Fantoni:2001}.
In a non-superfluid system, the calculation of the spin
susceptibility yields the $S^{-1}_\sigma$ sum-rule.

We calculate $S_\sigma^{0}$ by computing the spin--dependent pair 
correlation function and evaluating the structure function at $q=0$.
The spin correlation function is defined by
\begin{equation}
g_\sigma(r)=
\frac{1}{2\pi r^2 \rho N}
\sum_{i<j}\frac{\langle\psi\vert\delta(r_{ij}-r)
\boldsymbol\sigma_i\cdot\boldsymbol\sigma_j\vert\psi\rangle}
{\langle\psi\vert\psi\rangle} \,,
\end{equation}
where $\psi$ is the ground state of the system. The AFDMC method 
is useful to compute the expectation values of mixed operators 
like $\langle \Psi_T | O | \psi \rangle$.
We use Variational Monte Carlo (VMC) to extrapolate the value of 
operators that are given by $\langle O\rangle=2\langle O\rangle_{mix}
- \langle O\rangle_{vmc}$ as described in Ref. \cite{Gandolfi:2009,Gandolfi:2009b}.
The resulting $g(r)$ is used to obtain the structure function $S^0_\sigma(q)$.
We show $g_\sigma(r)$ and $S^0_\sigma(q)$ in Fig. \ref{fig:gofr}.
We finally evaluate $S_\sigma^0$ sum-rule by taking the
$q\rightarrow 0$ limit as indicated in Eq. (\ref{S2}).

The energy weighted-sum-rule can be calculated 
by the expectation value
of the tensor and spin-orbit interactions when $q=0$. 
For the Hamiltonian of Eq. (\ref{Ham})  we have
\begin{equation}
S_\sigma^{+1}=-\frac{4}{3N}\sum_{i<j}\left(3\langle v_3(r_{ij})S_{ij}\rangle+
\langle v_4(r_{ij})\boldsymbol L\cdot\boldsymbol S\rangle \right) .
\end{equation}

Because the variational wave function $\Psi_T$ used as input for AFDMC
contains neither tensor nor spin-orbit correlations, 
the most accurate
way to obtain these expectation values is by calculating the
energy as a function of the spin-orbit and tensor interaction strengths
and using the slope of the energy with respect to these couplings to
produce the true ground-state expectation values.

These initial calculations are performed with the AV8' NN interaction
without any three-nucleon interaction.  Based upon simple estimates
of the strength of the three-nucleon force, we would expect of order
$10-20 \%$ corrections to the sum-rules from the three-nucleon interaction.
We are exploring this dependence and will report these results
separately.

\begin{figure}[h]
\includegraphics[width=0.9\columnwidth]{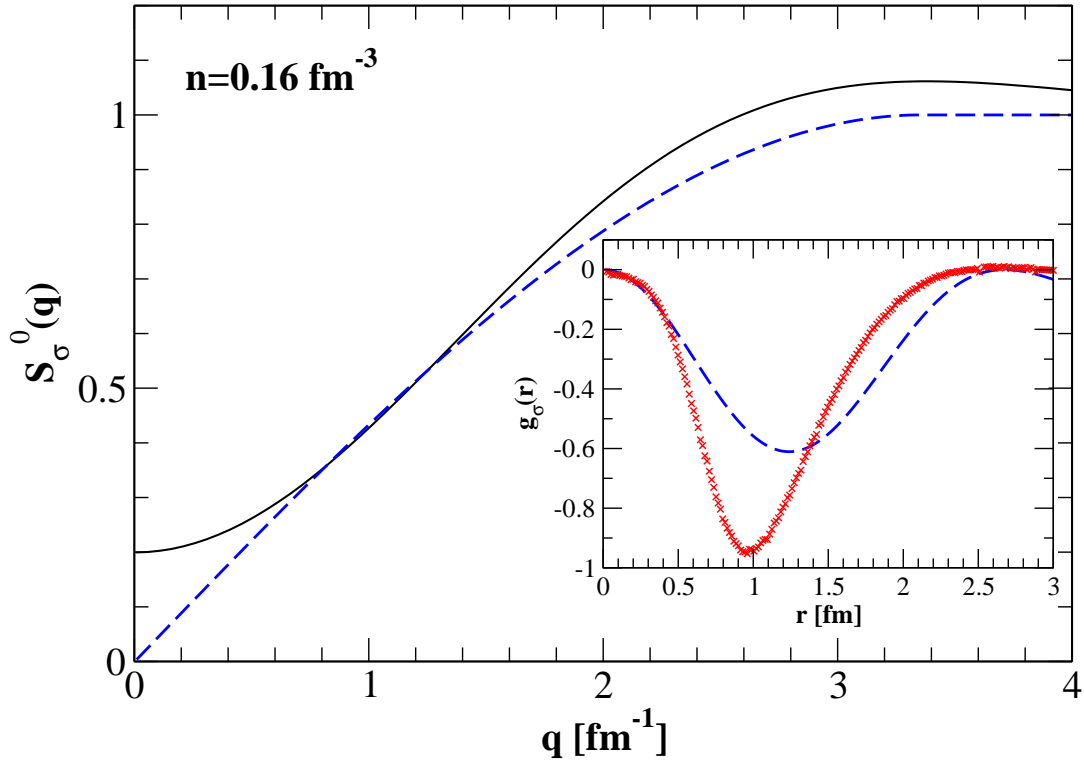}
\caption{(Color online) The static structure
function $S^0_\sigma(q)$  computed 
at saturation density. In the inset we show the corresponding spin pair correlation function $g_\sigma(r)$.
Free particle results are given as dashed lines.
\label{fig:gofr}}
\end{figure}

In computing the
ground state properties in AFDMC we neglect the role of pairing and
superfluidity.  This will restrict our study to the calculation of the neutrino
emissivity at temperatures that are large compared to the neutron pairing gaps
in neutron matter but still small compared to the Fermi energy. Thus, our
results will be applicable to ambient conditions in the supernova but will not
apply to old neutron stars where neutron matter is likely to be below the
superfluid critical temperature. For $T\ll \Delta$ where $\Delta\approx 1$ MeV is the superfluid gap, 
the number of quasi-particles is exponentially suppressed and response is vanishingly small. 
In vicinity of the critical temperature, Cooper pair breaking and formation, as well as collective 
modes  can enhance spin-fluctuations at a frequency $\omega \approx (1-2)\Delta$ \cite{Leinson:2006gh}. The spin response function and the neutrino emissivity in the superfluid phase is expected to be qualitatively different and is dominated by the pair recombination processes and the decay of finite energy collective modes \cite{Leinson:2010,Kolomeitsev:2011}. 
It may be possible in the future to examine this regime
more critically using techniques similar to those developed here.

The AFDMC results for the sum-rules are shown in Table \ref{tab:sum} where the
individual sum-rules and average excitation energies defined by $\bar{\omega}_0=
S_\sigma^{0} /S_\sigma^{-1}$ and $\bar{\omega}_1=   S_\sigma^{1} /S_\sigma^{0}$
are listed.  The density dependence of the $S_\sigma^0$ sum-rule is quite modest
over the range of densities considered.

% Requires the booktabs if the memoir class is not being used
\begin{table}[htbp]
   \centering
   \caption{AFDMC results for the sum-rules} % requires the topcapt package
   \begin{tabular}{@{} lccccc @{}} % Column formatting, @{} suppresses leading/trailing space
    \hline
      Density (fm$^{-3}$)    & $S_\sigma^{-1}$ (MeV$^{-1}$)    &  $S_\sigma^{0}$ & $S_\sigma^{+1}$ (MeV)& $\bar{\omega}_0$ (MeV) & $\bar{\omega}_1$ (MeV)\\ \hline
        $n=0.12$       & $0.0057(9) $  &  $0.20(1)$ & $8(1)$ & $ 35(9)$ & 40(8)  \\
           $n=0.16$  &  $0.0044(7)$ & $0.20(1)$ & $11(1)$ & 46(11) & 55(8) \\
           $n=0.20$  & 0.0038(6) & 0.18(1) & 14(1) & 47(12) & 78(10)   \\     
\hline
   \end{tabular}

      \label{tab:sum}
\end{table}

The spin susceptibilities shown in table \ref{tab:sum} correspond to
$\chi / \chi_F = 0.37, 0.34,$ and $0.34$ for $\rho = 0.12, 0.16$, and
0.20 fm$^{-3}$, where $\chi_F=mk_F/\pi^2$ is the spin susceptibility for free fermi gas.  At the lowest density this is very similar to results obtained in \cite{Fantoni:2001}, at the highest density our result is
approximately 20 per cent lower for the susceptibility.  The difference
may lie in the fact that the three-nucleon force used in \cite{Fantoni:2001}
is repulsive in unpolarized neutron matter, and less so in spin-polarized matter.

The average energies  $\bar{\omega}_0$ and $\bar{\omega}_1$ are extracted
from the sum-rules as 
estimates for the
energy of the peak of the response, and their difference is a measure
of the width of the distribution. 
The fact that the calculated $\bar{\omega}_0$ and $\bar{\omega}_1$ 
are fairly similar indicates a moderately narrow peak in the response.
A positive definite response requires $\bar{\omega}_1 \geq
\bar{\omega}_0$.  The peaks shift to higher energy with increasing density,
as expected. The tables also indicate that the strength 
distribution gets more diffuse with increasing density with 
strength being pushed out to higher energy.

\section{Asymptotic Form at High Energy}

In order to constrain the low-energy response relevant for astrophysical applications using sum-rules we need some knowledge of the behavior of $S_\sigma(\omega)$ at large $\omega$. In this regime the response probes the short time
behavior of the many-body correlation function and on general grounds we can
expect this to be dominated by two-particle dynamics. This intuitive expectation
can be cast in more formal terms using the operator product expansion originally
developed by Wilson as a standard technique in quantum field theory. The operator product expansion has been used to analyze short-time behavior of the density-density correlation function in a strongly
interacting non-relativistic fermi gas \cite{Braaten:2010,Son:2010}. Adapting this to the
spin-density operator, the relevant expansion in this case organizes
$S_\sigma(\omega)$ in terms of local operators in inverse powers $\omega$, and
is given by \beq
 \int dt ~e^{i\omega t}~\int d^3 x~
 \psi^\dagger \bm \sigma \psi (t,\bm R + \bm x) ~ \psi^\dagger \bm \sigma \psi(0,{\bm R- \bm x}) = i W_{1}(\omega) ~ {\cal O}^{(1)}(\bm R) + i W_{2}(\omega)~{\cal O}^{(2)}(\bm R)  + \cdots 
\eeq
where the expectation value of the local operators $ {\cal O}^{(n)}(\bm R)$
depends on the many-body ground state but the Wilson coefficients $W_{i}(\omega)$
depend only on few-body physics with $i$ incoming and outgoing asymptotic
states. For $\bm q=0$ the Wilson coefficient $W_{1}(\omega)$ vanishes
identically in spin saturated system and the leading contribution is due to
$W_{2}(\omega)$.  The functional form of $W_{2}(\omega)$ is determined by the
matrix elements of the spin operator between two-body scattering states. This implies that up to an over all constant which depends only on the ground state, $S_\sigma(\omega)$ at high frequency is determined by the the two-body matrix elements. In general, this will depend sensitively on the short-distance behavior of the two-nucleon interaction and will be model dependent. However, to extract the response at low energy in a model independent fashion it suffices to use in the two-body calculation, the same Hamiltonian employed in the calculation of the sum-rules in many-body calculation.

The spin response function $S(q,\omega)$ for two neutrons are evaluated as follows, 
\beq\label{sum2} S(q,\omega)\ =\ |<\psi_F|\hat{O}_A|\psi_I>|^2 \delta(\omega+E_I-E_F). \eeq 
For spin response at $q$=0, the operator is the sum of spins, $ \hat{O}_A\ =\vec{\sigma}_1+ \vec{\sigma}_2$. 
$\psi_I$ and $\psi_F$ are the eigenstates of two neutrons in spin-triplet states and take $\psi_I$ to be the ground state.

We have calculated these matrix elements using the same nuclear Hamiltonian employed in the AFDMC by solving the Schr\"odinger equation for two-neutrons with simple box boundary condition.
These results indicate that the high frequency behavior denoted as  $S^{\rm
high}_\sigma(\omega)$ is determined by two-body physics and has the following
asymptotic behavior 
\beq 
\label{eq:shigh}
S^{\rm high}_{\sigma} (\omega)  \simeq  \left(\frac{\omega_c}{\omega}\right)^i \,,
\eeq
where the density dependent quantity $\omega_c\simeq 100-150 $ MeV for the range of
densities considered here and for the nuclear interaction used we find that $i\approx 9$. 

As mentioned earlier the high frequency response will depend on model for nucleon-nucleon interactions at short-distance. For a correct description of the response at $\omega \ge 100$ MeV, the inclusion of two-body currents and explicit pion and $\Delta-$isobar degrees of freedom is likely to become important. However, since they are absent both in the many-body and two-body calculation, 
their consistent omission ensures that we can still obtain useful constraints on  $S_\sigma(\omega)$ 
at lower $\omega$ of interest without these ingredients. 

Using the two-body axial currents adjusted to reproduce measured tritium $\beta$ decay \cite{Schiavilla98}, we calculated the 
contributions to the static spin sum rule of Eq.~(\ref{sum2}) at $q=0$ due to the most important two-body currents -- the axial $\pi$-exchange $\Delta$-excitation current and $\pi$-exchange (pair) current. It was found to be a few percent of the total static spin sum rule. Therefore we expect the contribution of two body currents to the dynamic spin response function at zero momentum transfer to be around a few percent as well.

\section{ Low Energy Forms for the Response}
In the regime where neutron matter behaves like a Fermi liquid, the low-energy form of the response should be describable in terms of quasi-particles, though the coupling of the ground state to the quasiparticle pairs as well as the quasiparticle interactions may renormalize quantities in the calculated response. At $q=0$, a low-frequency form for $S_\sigma(\omega)$ has been computed in Refs.\cite{Raffelt:1996,Lykasov:2008} using the quasi-particle approximation and is given by  
\beq
\label{sqp}
S_\sigma(\omega)= \frac{N(0)}{n\pi}~\frac{\omega \tau_\sigma}{(1+G_0)^2+(\omega \tau_\sigma)^2}
\eeq
where the frequency dependent relaxation time $\tau_\sigma (\omega)$  is the
time-scale for damping of spin fluctuations, and $N(0)$ is the density of states
at the Fermi surface and $G_0$ is the Landau parameter that encodes the spin
susceptibility mentioned earlier Eq.~(\ref{S1}). This form of the response incorporates 
collisional broadening and mean field effects but it is mostly sensitive to the   
$\tau_\sigma (\omega)$. Note this type of functional form incorporates higher order terms in the scattering and thereby takes into account the Landau-Pomeranchuk-Migdal effect \cite{LPM1,LPM2,LPM3,Knoll96}. 

 The spin relaxation time
$\tau_\sigma$ is related to the quasi-particle scattering amplitude ${\cal
A}_{{\bm \sigma}_1,{\bm \sigma}_2}(\kk,\kk')$ and is given by
\cite{Lykasov:2008} 
\beqn\label{tau1} \frac{1}{\tau_\sigma(\omega+\epsilon_1)} \
&=& 2\pi\sum\limits_{2,3,4} 
|A|^2~{\cal F}~\delta(\omega+\epsilon_1+\epsilon_2-\epsilon_3-\epsilon_4)
\delta({\bm P}_1+{\bm P}_2-{\bm P}_3-{\bm P}_4). \\
\label{tau2} 
{\rm where}~  |A|^2 &=& \frac{1}{12}\sum\limits_{j}
{\rm Tr} \biggl[ \, {\cal A}_{{\bm \sigma}_1,{\bm \sigma}_2}(\kk,\kk') 
{\bm \sigma}_1^j \bigl[ ({\bm \sigma}_1 + {\bm \sigma}_2)^j \, , \,
{\cal A}_{{\bm \sigma}_1,{\bm \sigma}_2}(-\kk,\kk') \bigr] \biggr], 
\label{A2}
\eeqn 
%\begin{widetext}
%\beq\label{tau1} \frac{1}{\tau_\sigma} \ = 2\pi\sum\limits_{2,3,4} 
%|A|^2 [n_2(1-n_3)(1-n_4)+(1-n_2)n_3n_4]\delta(\omega+\epsilon_1+\epsilon_2-\epsilon_3-\epsilon_4)
%\delta({\bf P}_1+{\bf P}_2-{\bf P}_3-{\bf P}_4). \eeq 
%\end{widetext}
where the first sum is over the momenta of all initial states of particle 2 and
all final states of particle 3 and 4 and ${\cal F}= f_2(1-f_3)(1-f_4)+(1-f_2)f_3
f_4$ Pauli blocking factors where $f_i$ is the Fermi-Dirac distribution for
particle $i$ in the reaction $1+2 \rightarrow 3+4$ with incoming momenta 
$\bm P_1$ and $\bm P_2$ and outgoing momenta $\bm P_3$ and $\bm P_4$ and relative momenta 
$\kk={\bm P}_1-{\bm P}_3, \kk'={\bm P}_1-{\bm P}_4$. The squared matrix element in Eq.~(\ref{tau2}) is a sum over the spin projections 
$j=1,2,3$ of the spin operators  ${\bm \sigma}_1$ and ${\bm \sigma}_2$ acting on nucleons 1 and 2 respectively. In the limit of $\omega \ll E_F$ and $T\ll E_F$, the appropriate average
relaxation time from Eq.~(\ref{tau1}) reduces to 
\beq
\label{taulowk}
\frac{1}{\tau_\sigma}\ =\ C_\sigma~ \left[\left(\frac{\omega}{2\pi}\right)^2+T^2\right] 
\eeq
where $C_\sigma$ characterizes the strength of non-central interactions and
depends on the ambient density and $T$ is the temperature.  $C_\sigma$ has been
calculated using different models for the nucleon-nucleon potential for a range
of densities in Ref.~\cite{Bacca:2009}. At nuclear density they find that
$C_\sigma \simeq 0.22$ MeV$^{-1}$ for one pion exchange (OPE),  while it is reduced
to $C_\sigma \simeq 0.08$ MeV$^{-1}$ for realistic nucleon-nucleon and 
N$^3$LO $\chi$PT potentials. 

From Eq.~(\ref{taulowk}) it follows that the form of the 
quasi-particle approximation is valid
when $\omega  \ll (2\pi)^2/C_\sigma$ (and $\omega \ll E_F$) and inserting Eq.~(\ref{taulowk}) into
Eq.~(\ref{sqp}) we can obtain a low-frequency form of the zero temperature
structure function  \beq 
S^{\rm low}_\sigma(\omega, \bm q=0) = \frac{N(0)}{n\pi} \frac{\tilde{C}_\sigma~\omega}{1+ (1+G_0)^2(\tilde{C}_\sigma~\omega)^2}   
\label{eq:slow}
\eeq 
where $ \tilde{C}_\sigma=C_\sigma/(2\pi)^2$. This form of the structure function
satisfies the $S^{-1}_\sigma$ sum-rule by construction but produces divergent results
for the $S^{0}_\sigma$ and $S^{1}_\sigma$. In the following section we will combine the low
and high frequency forms in Eq.~(\ref{eq:slow}) and Eq.~(\ref{eq:shigh}), respectively,
with the sum-rule constraints discussed in the preceding section to construct a
structure function that can be used in calculations of the neutrino emissivity.   

Another commonly used limiting form of $S_\sigma(\omega)$ can be obtained by ignoring any many-particle correlations in the ground 
state. Here neutrons are distributed as free fermions in the ground state and excitations with $q=0$ and $\omega \neq 0 $
arise as two-particle two-hole states.  Following Ref.~\cite{Hanhart:2001} we denote this as the two-body (2b) response and this is given by  
\begin{equation}
\label{s2b}
S^{\rm 2b}_{\sigma}(\omega)=\frac{2}{3 \pi n } \int \left[\prod_{i=1..4}\frac{d^3p_i}{(2\pi)^3} 
\right]~(2\pi)^4 \delta^3({\bf P_1+P_2-P_3-P_4})~ \delta(\epsilon_1+\epsilon_2-\epsilon_3-\epsilon_4-\omega)~{\cal F}_2~{\cal H} \, ,
\label{eq:s2b}
\end{equation}
${\cal F}_4=f_1f_2(1-f_3)(1-f_4)$, and ${\cal H}$ is related to the square of the matrix element defined in Eq.~(\ref{A2}). For the $nn$ system where only the spin-triplet two-nucleon state contributes it is explicitly given by  
\beq
{\cal H}= \, \,
\frac{1}{\omega^2}~\sum_{\ms \ms'} \left| \langle 1 \ms',{\bf
p}'|\left[S,{\bf T}_{NN}\right]|{\bf p},1 \ms \rangle \right|^2,
\label{hii}
\eeq 
where $S$ is the total spin and ${\bf p} (${\bf p'}) is the relative  initial (final) momentum of the two-nucleon system. The matrix element is computed assuming plane-wave in and out states for the two-nucleon system and thus ignores many-body effects and initial and final state correlations. In Ref.~\cite{Hanhart:2001} it was argued that this result can be expected to be valid up to $\omega \simeq m_{\pi}$ in the absence of many-body corrections.

\section{Constructing $S_\sigma(\omega)$ } 
\label{sw}
While it is clear that a unique reconstruction of $S_\sigma(\omega)$ would
require an infinite number of moments, here we show that the lowest order sum-rules
and the the asymptotic forms discussed in the preceding section already provide
significant constraints. The three sum-rules $S^{-1}_\sigma,S^0_\sigma,S^{+1}_\sigma$ provide an useful
sampling of the function at low, intermediate and high energy respectively.
Its utility in constraining the neutrino emissivity will 
depend on the redistribution of strength due to finite temperature 
effects. We postpone a discussion of these finite temperature effects to the subsequent section. Here, using as guidance the limiting forms discussed previously, 
we examine simple ansatze for
the functional form of $S_\sigma(\omega)$ at $T=0$ by imposing sum-rule constraints. 
  
The striking feature of the sum-rule results shown in Table~\ref{tab:sum} is
that at nuclear density $\bar{\omega}_0 \approx \bar{\omega}_1$ and is comparable to the
Fermi energy $E_F=k_F^2/2m$. This suggests that the function $S_\sigma(\omega)$
contains significant strength in the region $\bar{\omega}_0$ to $\bar{\omega}_1$. To properly
account for this we study the following simple ansatz for the frequency
dependence of $S_\sigma(\omega)$ given by the form $S^{\rm low}_\sigma(\omega)$ but
the with a more complex behavior of $\tau(\omega)$ given by the following forms
\beqn
\frac{1}{\tau(\omega)} =  \left(\tilde{C_\sigma}\omega^2 + \alpha~ \frac{\omega^{2+n}}{(\omega + \omega_0)^2}\right)~\left(\frac{\omega_c}{\omega+\omega_c}\right)^m
\label{eq:tauinv}
\eeqn   
where the constants $\alpha,\omega_0$ and indices $n,m$ are fit to ensure that the three sum-rules and the asymptotic forms are satisfied. 
At low frequency, this ansatz ensures that out results coincide with the results obtained by in Ref.~\cite{Lykasov:2008} by Lykasov, Pethick and Schwenk where only the first term containing $\tilde{C_\sigma}$ contributes. On general grounds (unitarity of scattering amplitudes) at large frequency, pair excitation should be quenched due to the retarded nature of nuclear interaction, and this quenching is naturally incorporated through the asymptotic form discussed in relation to Eq.~(\ref{eq:shigh}).

To better understand the sensitivity of our results to the choice of parametrization,
we have also used a simple phenomenological form for the
spin response:
\begin{equation}
S_\sigma (\omega) \ = \ \alpha \frac{\omega^j}{(1 + (\omega/\omega_c)^i)^4}.
\label{eq:sphenom}
\end{equation}
The high frequency tail is forced to fall off appropriately by
choosing $4i-j = 9$.  The parameters $\alpha, \omega_c$, and $i$
are then fitted to the three sum-rules.  This simple form assures that
the response goes to zero at low frequency, has the correct high-frequency
tail, and has a single peak structure.  Comparisons of the two parametrizations
provide some information on the reliability of the extracted
spin response.

\begin{figure}[h]
\begin{center}
\includegraphics[width=0.9\columnwidth]{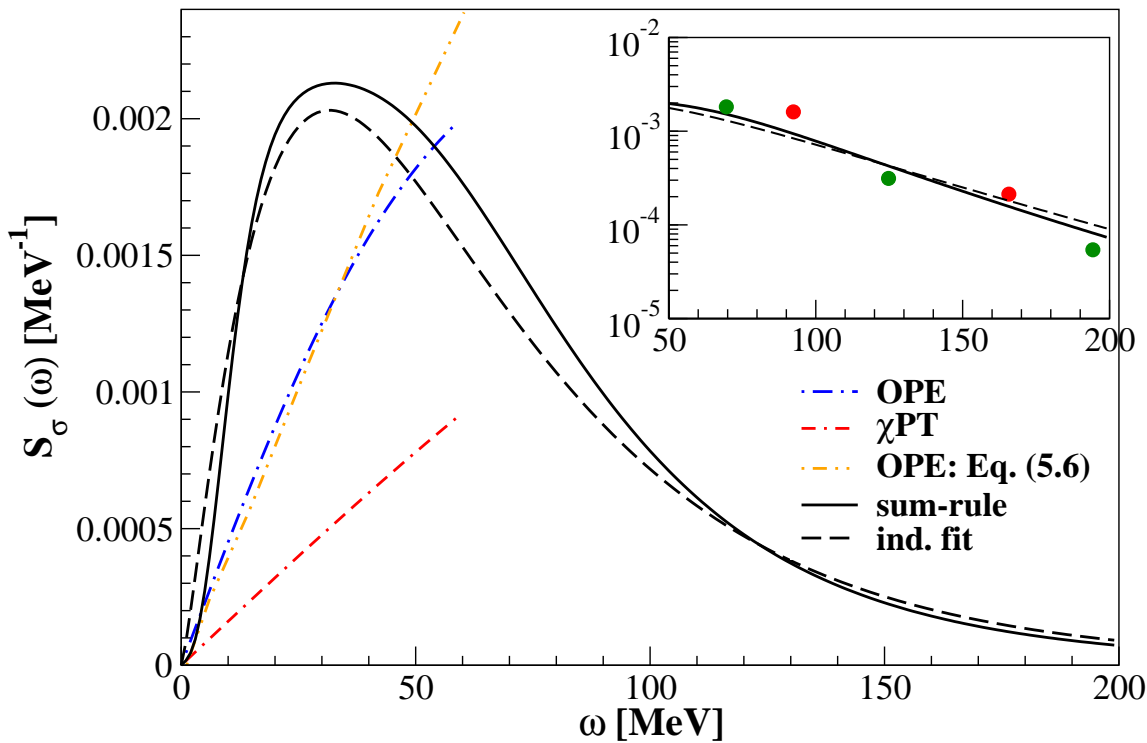}
\end{center}
\caption{(Color online) The spin response function $S_\sigma (q=0,\omega)$
of neutron matter at saturation density  obtained 
by fitting to AFDMC sum-rules
using two different ansatz are shown as the black solid and dashed curves. The inset compares the fits and the two-particle response at high
energy obtained by confining two neutrons in a spherical cavity of
radius 7 fm (red) or 8 fm (green). The linear, 
low-frequency forms predicted in Ref. \cite{Bacca:2009}, labeled as OPE and $\chi$PT are shown for comparison. The dot-dot-dashed curve is obtained using the two-body approach in Eq.~(\ref{s2b}) with OPE.} 
 \label{fig:2}
\end{figure}
 
Figure \ref{fig:2} shows the response function obtained by fitting
the sum-rules and the high-energy response at saturation density
using the two different parametrizations, Eqs. (\ref{eq:slow}) and
(\ref{eq:sphenom}).  For comparison, the low-frequency form of the structure function 
obtained in Ref.~\cite{Bacca:2009} are shown for the two choices of $\tilde{C_\sigma}$ corresponding to the OPE and $\chi$PT potentials discussed earlier. The form of the low-frequency response in Eq.~(\ref{sqp}) is valid only at $\omega \ll E_F$.  In the figure we also show the results from the two-body approach (described in Eq.~(\ref{s2b})) in the Born approximation with OPE.  At low frequency $\omega \le E_F/2$, it gives
similar results to the quasi-particle picture, then becomes larger at higher frequency since it includes the exact phase space integrals. The inset compares the fits and the two-particle response at high
energy obtained by confining two neutrons in a spherical cavity of
radius 7 fm (red) or 8 fm (green). 
The asymptotic forms and sum-rules force
significantly more strength at lower energy than obtained
previously.  

The simple phenomenological fit (dashed line - Eq.~(\ref{eq:sphenom}) and
the fit to the quasi-particle form 
(solid line - Eqs. (\ref{eq:slow}) and (\ref{eq:tauinv})) produce very similar response
functions. In addition to the sum-rule constraints, we are
forcing the response to go to zero at low frequency, have a single
peak structure, and to fall off fairly rapidly at high-energy
as obtained from the two-neutron response.  Combined, 
these considerations place fairly tight constraints on the spin
response of neutron matter.

\begin{figure}[h]
\begin{center}
\includegraphics[width=0.8\columnwidth]{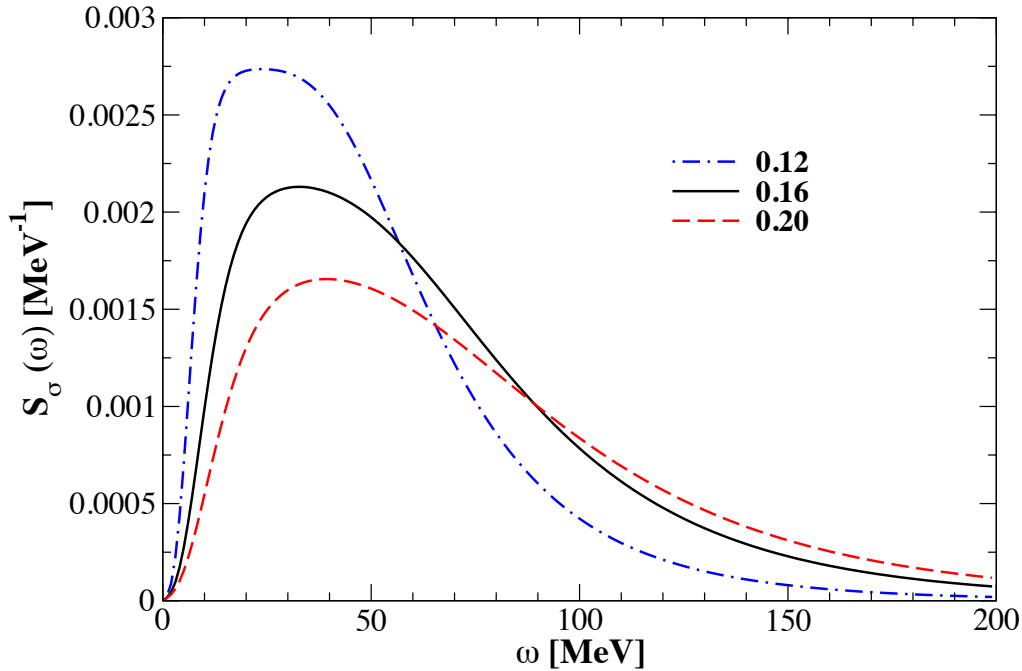}
\end{center}
\caption{(Color online) The spin response function $S_\sigma (q=0,\omega)$
of neutron matter at $\rho = 0.12, 0.16,$ and $0.20$ fm$^{-3}$ from fits to
AFDMC sum-rules results at zero temperature .
}
\label{fig:fitsden}
\end{figure}

In Figure \ref{fig:fitsden} we compare the response functions obtained
over a range of densities $\rho = 0.12, 0.16$ and $0.20$ fm$^{-3}$.
As expected from the sum-rules, the peak of the response functions
shifts to larger energy with increasing density. The tensor and
spin-orbit correlations are naturally of shorter range at the higher
densities where the mean inter-particle spacing is shorter, and hence
the peak shifts to higher energy.  
The total strength in the response is fairly flat over the regime
of densities we consider as obtained in the sum-rule calculations for $S_0$.  

Finally,
at higher density the distribution is somewhat broader as $\omega_1$ increases more rapidly 
with density than $\omega_0$. Both $\omega_0$ and $\omega_1$
increase rapidly, presumably associated with
the increasing importance of the shorter-range components of the
nuclear force at and above saturation density. While we expect this trend to be qualitatively correct 
contributions due to three-body forces and from two-body currents are able to play a role in modifying this 
behavior.

\section{Extension to finite temperature and impact on neutrino production}

The AFDMC method we employ is restricted to zero temperature and we have not 
explicitly computed the temperature corrections to the sum-rules. Hence there will be several 
caveats to consider when using our results in finite temperature applications 
such as supernova where $S_\sigma(\omega)$ plays a role in neutrino production rates. 
To discus these we first note that there are three fundamental energy scales inherent to our 
present analysis of the structure function and the neutrino emissivity. They are: (i) typical
energy at which the structure function has significant strength and is given by 
$\bar{\omega}_0$ and $\bar{\omega}_1$; (ii) the energy scale at which the structure
function is sampled in the neutrino emissivity and is denoted as $ \omega_\nu$  
from the expression for $Q$ in Eq.~(\ref{eloss}) we expect that $ \omega_\nu
\simeq 5-6 ~T$ ; and (iii) the high energy scale $\omega_c$ at which the asymptotic two-body behavior
dominates. 

At very low temperature where $\omega_\nu \ll \bar{\omega}_0$ and
$\omega_\nu \ll  \bar{\omega}_1$, the sum-rules do not provide useful constraints.
Here the low-frequency form of  $S_\sigma(\omega)$ given in Eq.~(\ref{eq:slow})  can
be used to calculate the neutrino emissivity with the requirement that $\omega
_\nu \ll  E_F$ and $\tilde{C}_\sigma \omega_\nu \ll1$. In practice, at nuclear
density, the condition that $\omega _\nu \ll  E_F$ is more restrictive and
limits the use of the low frequency form to region where $ T \le E_F/6$. At
intermediate temperature when  $\omega_\nu \simeq  \omega_0$ or $\omega_1$ and $ T \le E_F$, 
the zero temperature sum-rule constraints on the form of $S^0_\sigma$ and $S^1_\sigma$ become relevant. 
Here the temperature is intermediate and corrections to the $T=0$ sum-rules are expected to be small due to 
the Pauli principle.    

At finite temperature, the dynamic structure factor obey detailed-balance 
\begin{equation} 
S_\sigma(-\omega) = \exp{\left(-\frac{\omega}{T}\right)}~ S_\sigma(\omega) \,, 
\end{equation} 
and this is reflected in Eq.~(\ref{eloss}) where the neutrino emissivity where the exponential term accounts for the fact that 
neutrino emission corresponds thermal fluctuations in which $\omega$ is negative.  There are residual temperature dependencies 
in the function $S_\sigma(\omega)$.  First, from the fluctuation-dissipation theorem we have 
\begin{equation} 
S_\sigma(\omega) = -2~(1-\exp{ \left(-\frac{\omega}{T}\right)} )^{-1} ~{\rm Im}~\Pi^{\rm R} (\omega) \,, 
\end{equation} 
where $\Pi^{\rm R}$ is the retarded polarization function which is an odd function of $\omega$ and vanishes at $\omega=0$. To extend to finite temperature the zero-temperature ansatze in \S\ref{sw} need to be multiplied the factor  
$(1-\exp{ \left(-\omega/T\right)} )^{-1}$. A second source of temperature corrections  arise from the fact that at low frequency the spin relaxation time $\tau_\sigma^{-1} \simeq  C_\sigma T^2$ is dominated by scattering between thermally excited quasi-particles as described in Eq.~(\ref{eq:tauinv}). We incorporate this expected behavior by using 
the finite temperature expression for $\tau_\sigma$ given in Eq.~(\ref{eq:tauinv}) in the low-frequency form given 
for $S_\sigma(\omega)$ in Eq.~(\ref{eq:slow}).  Other sources of temperature corrections exist such as those arising from transitions in which the excited many-particle states does not decay to the ground state and should be investigated in the future. We leave this for future work as it would require the development of finite temperature QMC techniques.

With the aforementioned finite temperature extensions we employ the $S_\sigma(\omega)$ obtained using the sum-rule constraints 
to compute the neutrino emissivity. In Figure \ref{fig:nmfp}, the resulting energy loss rate $Q$ are shown for various
temperatures $T$. The large strength required by the sum-rules at intermediate energy  leads to a larger neutrino emissivity compared to the simple extrapolation of results obtained in the quasi-particle approximation with only two-particle two-hole excitations.  Our results are almost a factor of 2 larger than those obtained using either $\chi$PT in Ref.~\cite{Bacca:2009} or those obtained directly from nucleon-nucleon phase-shifts in Ref.~ \cite{Hanhart:2001} at $T \le$ 5 MeV. We suspect that this enhancement is due to correlations in the ground state that are not captured in the quasi-particle approximation. 

\begin{figure}[th]
\begin{center}
\includegraphics[width=0.9\columnwidth]{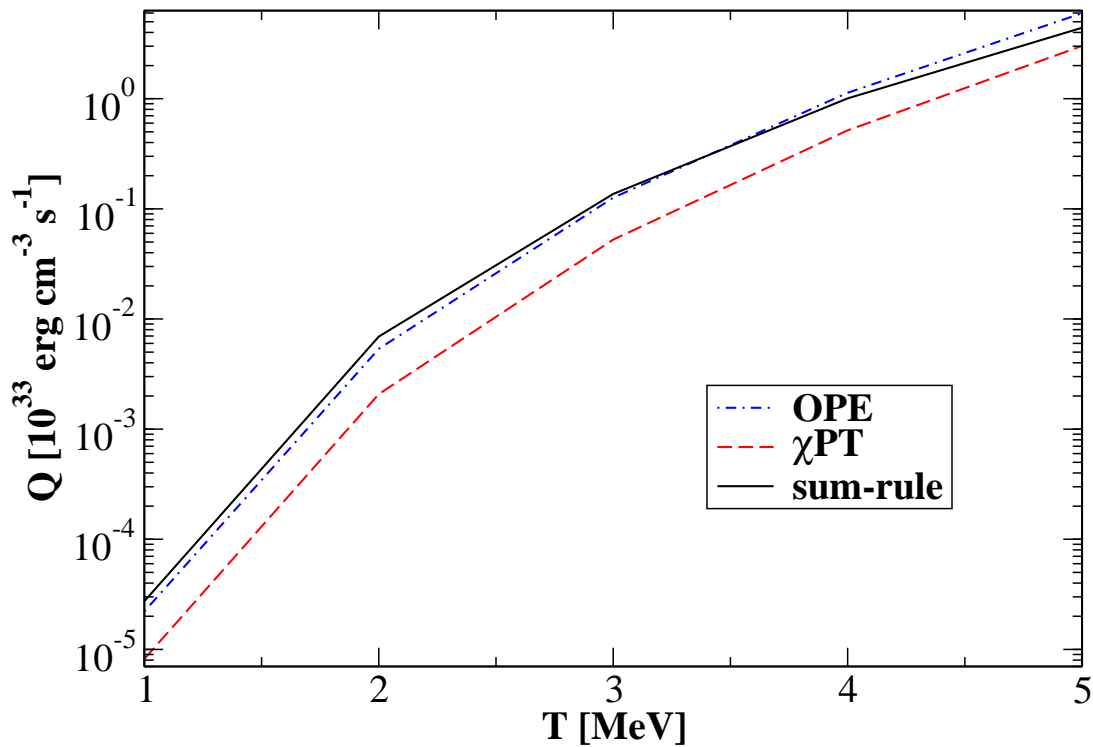}
\end{center}
\caption{(Color online) The energy-loss rate $Q$ at various
temperature as defied in Eq.~(\ref{eloss}), for OPE, $\chi$PT and our
results. \label{fig:nmfp}}
\end{figure}

\section{Discussion} 

Our study of the zero-temperature sum-rules suggests that the spin 
response function of neutron matter has a significant strength at 
energy between $40-60$ MeV in the vicinity of nuclear density. This strength 
should be accessible at temperatures of the order of $5-30$ MeV encountered 
in the supernova environment and could influence the rate of neutrino pair 
production from nucleon-nucleon processes. Although the zero-temperature sum-rules 
do not directly constrain the finite temperature response functions 
needed in the calculation of the neutrino emissivity they provide a 
useful guidance. For example, they can be used to test predictions obtained using 
quasi-particle methods at zero temperature, and with some caveats can be extended to 
low temperature where response is still dominated by transitions 
to the ground state. 

Comparisons with earlier calculations of the dynamic structure 
factor indicate they significantly under-predict the response 
in the regime where $\omega \simeq E_F/5-E_F$ for the densities considered.   There are several possible resolutions to this discrepancy.
The quasi-particle interactions and dispersion relations used in earlier studies may not be adequate as finite density effects are ignored. Similarly the use of plane wave states augmented with the
T-matrix, may be too simple to reproduce the coupling between the ground state
and the excitations.  Alternatively, our ansatze for the response function 
may be too simple to capture the complex structure of $S_\sigma(\omega)$. 

All of these possibilities can be 
studied in more detail. To quantify the interplay between increasing 
phase space and decreasing strength of the two-body interaction with increasing $\omega$ we have calculated $S_\sigma(\omega)$ in the standard 
approach using realistic potentials.  Within the QMC approach 
there are two ways to address these issues. First, the calculation 
of higher order moments at zero temperature can provide additional 
constraints and test our ansatze at intermediate energy. Second, extensions to 
finite temperature will shed light on the importance of transitions 
not involving the ground state. We hope to pursue these in future work.        
\section{Acknowledgements}

We thank Chris Pethick, Daniel Phillips, Kevin Schmidt, and Achim Schwenk for helpful
discussions.  This work was supported in part by a grant from
the DOE under contracts DE-AC52-06NA25396, DE-FG02-00ER41132 and 
collaborations were facilitated by DOE funds for the topical collaboration to study ``Neutrinos and Nucleosynthesis in Hot and Dense Matter". This work is also supported by the LDRD program at Los Alamos National
Laboratory (LANL). Computations for this work were carried out through
Open Supercomputing at LANL, and at the National Energy Research Science
Computing (NERSC).

\bibliographystyle{apsrev}
\bibliography{biblio}

\end{document}